\documentclass[conference]{IEEEtran}
\IEEEoverridecommandlockouts
\usepackage{cite}
\usepackage{amsmath,amssymb,amsfonts}
\usepackage{algorithmic}
\usepackage{graphicx}
\usepackage{textcomp}
\usepackage{xcolor}
\usepackage{comment}

\usepackage{babel}
\usepackage{graphicx}
\usepackage{adjustbox}
\usepackage{float}

\usepackage[font=small]{caption}
\usepackage[mode=tex]{standalone} 
\usepackage{amsmath,amssymb,amsfonts,bbm,bm,mathrsfs}
\usepackage{color}
\usepackage{algorithmic}
\usepackage{graphicx}
\usepackage{textcomp}
\usepackage{xcolor}
\usepackage{amsthm}

\usepackage[ruled,vlined,linesnumbered]{algorithm2e}

\usepackage[mode=tex]{standalone} 

\usepackage{tikz}

\usepackage{chronosys}
\usepackage{xspace}
\usetikzlibrary{shapes,arrows,calc,positioning,fit,automata}
\usetikzlibrary{decorations.markings}
\usetikzlibrary{overlay-beamer-styles}
\usepackage{fontawesome}

\usepackage{pgfplots}
\usepackage{pgfplotstable}
\usepackage{filecontents}

\usepackage{caption,subcaption}

\usepackage{soul}

\definecolor{green}{rgb}{0, 0.5, 0}
\definecolor{pink}{rgb}{1, 0, 1}

\newcommand{\ozan}[1]{{\color{blue}#1}}

\DeclareMathOperator*{\argmax}{arg\,max}

\def\BibTeX{{\rm B\kern-.05em{\sc i\kern-.025em b}\kern-.08em
    T\kern-.1667em\lower.7ex\hbox{E}\kern-.125emX}}

\begin{document}

\title{Neural Compress-and-Forward for the Primitive Diamond Relay Channel\\
\thanks{This material is based upon work supported in part by NYU Wireless, the National Science Foundation under grant no. 21482148315, and funds from federal agency and industry partners as specified in the Resilient \& Intelligent NextG Systems (RINGS) program.}
}

\author{
    \IEEEauthorblockN{
        Ozan Ayg{\"u}n,
        Ezgi~{\"O}zyılkan,
        Elza Erkip
    }
    \IEEEauthorblockA{
        Department of Electrical and Computer Engineering, New York University, Brooklyn, NY \\
        \texttt{\{ozan, ezgi.ozyilkan, elza\}@nyu.edu}
    }
    \vspace{-3em}
}

\maketitle

\begin{abstract}

The diamond relay channel, where a source communicates with a destination via two parallel relays, is one of the canonical models for cooperative communications. We focus on the primitive variant, where each relay observes a noisy version of the source signal and forwards a compressed description over an orthogonal, noiseless, finite-rate link to the destination. Compress-and-forward (CF) is particularly effective in this setting, especially under oblivious relaying where relays lack access to the source codebook. While neural CF methods have been studied in single-relay channels, extending them to the two-relay case is non-trivial, as it requires fully distributed compression without any inter-relay coordination. We demonstrate that learning-based quantizers at the relays can harness input correlations by operating remote, yet in a collaborative fashion, enabling effective distributed compression in line with Berger--Tung-style coding. Each relay separately compresses its observation using a one-shot learned quantizer, and the destination jointly decodes the source message. Simulation results show that the proposed scheme, trained end-to-end with finite-order modulation, operates close to the known theoretical bounds. These results demonstrate that neural CF can scale to multi-relay systems while maintaining both performance and interpretability.

\end{abstract}

\begin{IEEEkeywords}
diamond relay channel, compress-and-forward, distributed compression, task-aware compression, binning.
\end{IEEEkeywords}

\section{Introduction}

Modern wireless systems, including cellular and cell-free architectures, increasingly rely on distributed infrastructures where remote radio heads handle radio and front-end processing, while a central unit performs decoding and coordination~\cite{sanderovich2008communication}. Distributed cooperative relaying is the basic element in what is known as the Cloud Radio Access Network (CRAN), where there are several relays, each of which possesses a capacity-constrained backhaul link to a central unit~\cite{park2014fronthaul, aguerri2019capacity}, also referred to as a cloud decoder. Motivated by CRAN, in this paper, we study the \textit{diamond relay channel} (DRC), a canonical model consisting of a source, two relays, and a destination, where the relays assist in transmission via two separate links to the destination, and no direct link exists between the source and the destination~\cite{schein2001distributed}. 

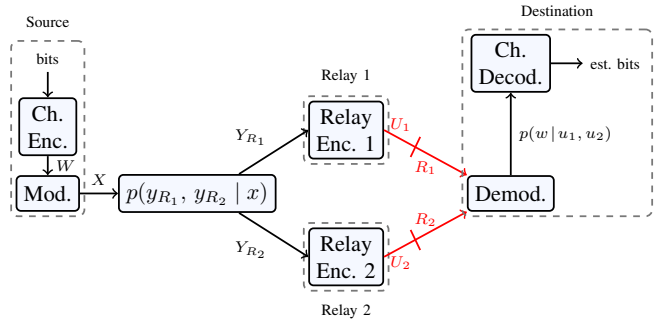
\begin{figure}
    \centering
    \definecolor{linkred}{RGB}{200,38,38}
\definecolor{boxfill}{RGB}{245,248,255}
\definecolor{framegray}{RGB}{120,120,120}

\begin{tikzpicture}[
    every node/.style={font=\footnotesize},
    blk/.style={draw,thick,rounded corners=2pt,
                minimum width=0.9cm,minimum height=0.55cm,
                fill=boxfill,align=center,font=\normalsize},
    frame/.style={draw=framegray,dashed,rounded corners=3pt,thick,inner sep=2pt},
    lab/.style={font=\scriptsize},
    redlink/.style={draw=linkred,red,thick,->},
    data/.style={draw,thick,->},
]

\def\relayShift{10mm}

\coordinate (relayMid) at (0, 0);
\coordinate (relayUp) at ($(relayMid)+(0,\relayShift)$);
\coordinate (relayDown) at ($(relayMid)-(0,\relayShift)$);

\node[blk] (ENC) at ($(relayMid)+(-7.5cm,0)$) {Ch.\\Enc.};
\node[lab,above=4mm of ENC] (BITS) {bits};
\draw[data] (BITS) -- (ENC);

\node[blk,below=3mm of ENC] (MOD) {Mod.};
\draw[data] (ENC) -- node[lab,right] {$W$} (MOD);

\node[frame,fit=(BITS)(ENC)(MOD)] (SRCF) {};
\node[lab,above=2pt of SRCF.north] {Source};

\node[blk,right=6mm of MOD] (CHAN) {$p(y_{R_1},\,y_{R_2}\mid x)$};
\draw[data] (MOD.east) -- node[lab,above] {$X$} (CHAN.west);

\node[blk,right=5mm of CHAN, yshift=+\relayShift] (ENC1) {Relay\\Enc.\ 1};
\node[frame,fit=(ENC1)] (R1F) {};
\node[lab,above=2pt of R1F.north] {Relay 1};

\node[blk,right=5mm of CHAN, yshift=-\relayShift] (ENC2) {Relay\\Enc.\ 2};
\node[frame,fit=(ENC2)] (R2F) {};
\node[lab,below=2pt of R2F.south] {Relay 2};

\draw[data] (CHAN.25)  -- node[lab,above left=-1pt] {$Y_{R_1}$} (ENC1.west);
\draw[data] (CHAN.-25) -- node[lab,below left=-1pt] {$Y_{R_2}$} (ENC2.west);

\node[blk,right=30mm of CHAN] (DEM) {Demod.};

\draw[redlink]
  (ENC1.east) -- node[lab,above,pos=0.2] {$U_1$}
               node[lab,below,pos=0.5] {$R_1$}
               (DEM.north west);

\draw[redlink]
  (ENC2.east) -- node[lab,below,pos=0.2] {$U_2$}
               node[lab,above,pos=0.5] {$R_2$}
               (DEM.south west);

\draw[red,thick]
  ($ (ENC1.east)!0.25!(DEM.north) + (2pt,2pt) $) -- ($ (ENC1.east)!0.35!(DEM.north) + (-8pt,-5pt) $);
\draw[red,thick]
  ($ (ENC2.east)!0.25!(DEM.south) + (2pt,-2pt) $) -- ($ (ENC2.east)!0.35!(DEM.south) + (-8pt,5pt) $);

\node[blk,above=13mm of DEM] (DEC) {Ch.\\Decod.};
\draw[data] (DEM.north) -- node[lab,right] {$p(w\!\mid\! u_1,u_2)$} (DEC.south);

\node[lab,right=5mm of DEC] (ESTBITS) {est. bits};
\draw[data] (DEC.east) -- (ESTBITS);

\node[frame,fit=(DEM)(DEC)(ESTBITS)] (DESTF) {};
\node[lab,above=2pt of DESTF.north] {Destination};

\end{tikzpicture}
    \vspace{-1.5em}
    \caption{Primitive diamond relay channel model under consideration. Red links indicate orthogonal (or out-of-band) relaying between the two relays and the destination.
    }
    \label{fig:system_model}
\end{figure} 

When relay-to-destination links are rate-limited, efficient compression becomes essential for maintaining high throughput~\cite{park2014fronthaul}. The \textit{primitive} DRC, where each relay forwards its noisy observation over an orthogonal (or out-of-band) finite-rate link, provides a useful abstraction~\cite{kang2015gaussian}. In this model, the compress-and-forward (CF) strategy~\cite{Cover_Gamal} is particularly effective, especially under \textit{oblivious relaying}, where the relays are unaware of the source codebook~\cite{simeone2011codebook, katz2019gaussian, katz2024gaussian}. The oblivious setting aligns naturally with learning-based designs, where relays learn to compress their observations directly from data without requiring any knowledge of the transmission strategy adopted by the source.

Motivated by this connection, we extend prior work on neural CF for single-relay channels to the primitive DRC with independent Gaussian noises~\cite{ozyilkan2024learning, ozyilkan2024spawc}. This extension is non-trivial, as it requires fully distributed compression without direct communication between relays. In this setting, the information-theoretic technique known as \textit{compress–bin}~\cite{network_info_theo} offers a good strategy, yet it remains challenging to implement in practice. To the best of our knowledge to date, there are no existing practical CF schemes that perform distributed compress–bin strategies considering multiple relays. 

Here, we propose an end-to-end learned framework where each relay separately compresses its observation using a one-shot neural quantizer, and the destination decodes the source message from the relays. Our contributions are as follows: 
\begin{itemize}
    \item The learned compressors recover \textit{binning} behavior consistent with prior work regarding Berger--Tung-style distributed compression~\cite[Chapter~12]{network_info_theo} without imposing an explicit structure onto the quantizer, enabling near-optimal performance under stringent rate constraints. 
    \item Simulation results show that the proposed scheme, trained end-to-end with finite-order modulation, operates close to the theoretical bounds for the Gaussian primitive diamond channel. These results pave the way for scalable and interpretable neural CF adopted in multi-relay systems.
    \item We extend our findings on different modulation schemes such as real-valued BPSK, 4-PAM, and 8-PAM, as well as complex-valued 4-QAM and 16-QAM, and demonstrate that the effectiveness of our learned approach applies to higher modulation orders as well.
\end{itemize}

\section{System Model} \label{sec:asilomar-system-model}

We consider the primitive DRC model in~\cite{ayfer2019new}, as illustrated in Fig.~\ref{fig:system_model}, where we consider a finite-order modulation in which an index $W \in \{1, \ldots, |\mathcal{X}|\}$ is mapped to a symbol $X \in \mathcal{X}$, with $\mathcal{X} \subset \mathbb{R}$ denoting a constellation of cardinality $|\mathcal{X}|$.  The Gaussian primitive DRC is given by:
\begin{align} \small
    Y_{R_{1}} &= X + N_{R_{1}}, & Y_{R_{2}} &= X + N_{R_{2}},
\end{align}
where $X$ is the transmitted signal from the source with power constraint $P$, and $Y_{R_1}$, $Y_{R_2}$ are the signals received at Relay 1 and 2, respectively. The noises at both relays $N_{R_1} \sim \mathcal{N}(0, \sigma_{R_1}^2)$ and $N_{R_2} \sim \mathcal{N}(0, \sigma_{R_2}^2)$ are independent.



We consider a DRC with the relays operating in an oblivious setting, meaning they do not have access to the codebook shared between the source and the destination \cite{simeone2011codebook}. The relay-to-destination links are orthogonal and noiseless, with capacities $R_1$ and $R_2$ bits per channel use, respectively.

\section{Neural Compress-and-Forward Schemes for the Primitive Diamond Relay Channel}

Following the prior neural CF framework~\cite{ozyilkan2024learning, ozyilkan2024spawc}, the goal is to jointly learn the quantizers at the relays, which map the relay observations $Y_{R_1}$ and $Y_{R_2}$ to compressed descriptions $U_1$ and $U_2$, respectively. As part of the training setup, a soft demodulator at the destination is also learned, generating a distribution on the coded symbols $W$ based on the received compressed signals.

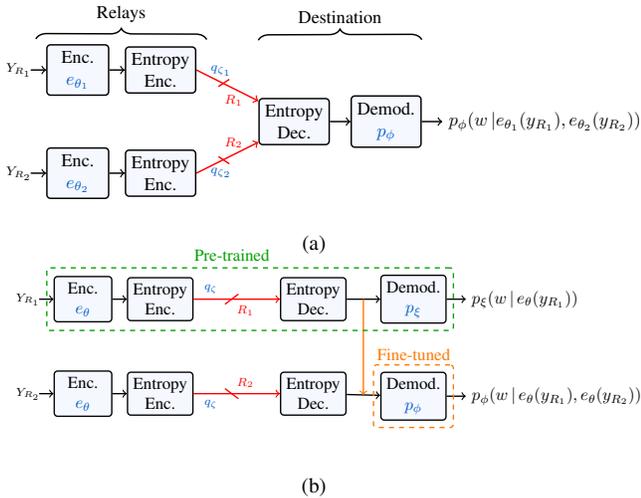
\begin{figure}
    \centering
    \begin{subfigure}[b]{\columnwidth}
      \definecolor{linkred}{RGB}{200,38,38}
\definecolor{paramblue}{RGB}{0,92,200}
\definecolor{boxfill}{RGB}{245,248,255}

\begin{tikzpicture}[
    every node/.style={font=\footnotesize},
    blk/.style={draw,thick,rounded corners=2pt,
                minimum width=1.2cm,minimum height=0.6cm,
                fill=boxfill,align=center,font=\normalsize},
    lab/.style={font=\scriptsize},
    redlink/.style={draw=linkred,red,thick,->},
    data/.style={draw,thick,->},
    blueL/.style={text=paramblue},
]

\node[blk] (ENC1) at (0, 0.9) {Enc.\\\textcolor{paramblue}{$e_{\theta_1}$}};
\node[blk,right=3mm of ENC1] (ENT1) {Entropy\\Enc.};

\node[blk,below=10mm of ENC1] (ENC2) {Enc.\\\textcolor{paramblue}{$e_{\theta_2}$}};
\node[blk,right=3mm of ENC2] (ENT2) {Entropy\\Enc.};

\node[blk,right=19mm of $(ENT1)!0.5!(ENT2)$] (DEC) {Entropy\\Dec.};
\node[blk,right=4mm of DEC] (DEM) {Demod.\\\textcolor{paramblue}{$p_{\phi}$}};

\node[align=left,right=4mm of DEM] (PRED) {\normalsize$p_{\phi}(w\,|e_{\theta_1}(y_{R_1}),e_{\theta_2}(y_{R_2}))$};

\draw[data] ($(ENC1.west)+(-3mm,0)$) -- node[lab,left] {$Y_{R_1}$} (ENC1.west);
\draw[data] ($(ENC2.west)+(-3mm,0)$) -- node[lab,left] {$Y_{R_2}$} (ENC2.west);

\draw[data] (ENC1) -- (ENT1);
\draw[data] (ENC2) -- (ENT2);

\draw[redlink] (ENT1.east) -- node[lab,above,pos=0.4] {\textcolor{paramblue}{$q_{\zeta_1}$}} node[lab,below,pos=0.6] {\textcolor{red}{$R_1$}} (DEC.150);
\draw[redlink] (ENT2.east) -- node[lab,below,pos=0.4] {\textcolor{paramblue}{$q_{\zeta_2}$}} node[lab,above,pos=0.6] {\textcolor{red}{$R_2$}} (DEC.210);

\draw[red,thick]
  ($ (ENT1.east)!0.4!(DEC.130) + (1pt,1pt) $) -- ($ (ENT1.east)!0.4!(DEC.130) + (-5pt,-3pt) $);
\draw[red,thick]
  ($ (ENT2.east)!0.4!(DEC.230) + (1pt,-1pt) $) -- ($ (ENT2.east)!0.4!(DEC.230) + (-5pt,3pt) $);

\draw[data] (DEC) -- (DEM);
\draw[data] (DEM) -- (PRED);

\draw[decorate,decoration={brace,amplitude=6pt},thick] ($(ENC1.north west)+(-2mm,1mm)$) -- ($(ENT1.north east)+(2mm,1.3mm)$) node[midway,above=2mm] {\normalsize Relays};

\draw[decorate,decoration={brace,amplitude=6pt},thick] ($(DEC.north west)+(0,11.2mm)$) -- ($(DEM.north east)+(0,10.8mm)$) node[midway,above=2mm] {\normalsize Destination};

\end{tikzpicture}
      \vspace{-0.5em}
       \caption{}
        \label{fig:dist_model} 
    \end{subfigure}
    \vspace{-0.5em}
    \begin{subfigure}[b]{\columnwidth}
    \vspace{-0.5em}
      \definecolor{linkred}{RGB}{200,38,38}
\definecolor{paramblue}{RGB}{0,92,200}
\definecolor{boxfill}{RGB}{245,248,255}
\definecolor{pretrain}{RGB}{0,160,0}
\definecolor{finetune}{RGB}{255,120,0}

\begin{tikzpicture}[
    every node/.style={font=\footnotesize},
    blk/.style={draw,thick,rounded corners=2pt,
                minimum width=1.2cm,minimum height=0.6cm,
                fill=boxfill,align=center,font=\normalsize},
    lab/.style={font=\scriptsize},
    redlink/.style={draw=linkred,red,thick,->},
    data/.style={draw,thick,->},
    orangearrow/.style={draw=finetune,thick,->},
    blueL/.style={text=paramblue},
    prebox/.style={draw=pretrain,dashed,thick,rounded corners=2pt,inner sep=4pt},
    finetune/.style={draw=finetune,dashed,thick,rounded corners=2pt,inner sep=4pt}
]

\node[blk] (ENC1) at (0, 0.9) {Enc.\\\textcolor{paramblue}{$e_{\theta}$}};
\node[blk,right=3mm of ENC1] (ENT1) {Entropy\\Enc.};

\node[blk,below=10mm of ENC1] (ENC2) {Enc.\\\textcolor{paramblue}{$e_{\theta}$}};
\node[blk,right=3mm of ENC2] (ENT2) {Entropy\\Enc.};

\node[blk,right=18mm of ENT1] (DEC1) {Entropy\\Dec.};
\node[blk,right=18mm of ENT2] (DEC2) {Entropy\\Dec.};
\node[blk,right=7mm of DEC1] (DEM1) {Demod.\\\textcolor{paramblue}{$p_{\xi}$}};
\node[align=left,right=4mm of DEM1] (PRED1) {\normalsize$p_{\xi}(w\,|\,e_{\theta}(y_{R_1}))$};

\node[blk,below=10mm of DEM1] (DEM2) {Demod.\\\textcolor{paramblue}{$p_{\phi}$}};
\node[align=left,right=4mm of DEM2] (PRED2) {\normalsize$p_{\phi}(w\,|\,e_{\theta}(y_{R_1}),e_{\theta}(y_{R_2}))$};

\draw[data] ($(ENC1.west)+(-3mm,0)$) -- node[lab,left] {$Y_{R_1}$} (ENC1.west);
\draw[data] ($(ENC2.west)+(-3mm,0)$) -- node[lab,left] {$Y_{R_2}$} (ENC2.west);

\draw[data] (ENC1) -- (ENT1);
\draw[data] (ENC2) -- (ENT2);

\draw[redlink] (ENT1.east) -- node[lab,above,pos=0.2] {\textcolor{paramblue}{$q_{\zeta}$}} node[lab,below,pos=0.6] {\textcolor{red}{$R_1$}} (DEC1.west);
\draw[redlink] (ENT2.east) -- node[lab,below,pos=0.2] {\textcolor{paramblue}{$q_{\zeta}$}} node[lab,above,pos=0.6] {\textcolor{red}{$R_2$}} (DEC2.west);

\draw[red,thick]
  ($ (ENT1.east)!0.5!(DEC1.west) + (1pt,3pt) $) -- ($ (ENT1.east)!0.5!(DEC1.west) + (-7pt,-3pt) $);
\draw[red,thick]
  ($ (ENT2.east)!0.5!(DEC2.west) + (1pt,-3pt) $) -- ($ (ENT2.east)!0.5!(DEC2.west) + (-7pt,3pt) $);

\draw[data] (DEC1) -- (DEM1);
\draw[data] (DEC2) -- (DEM2);
\draw[data] (DEM1) -- (PRED1);
\draw[data] (DEM2) -- (PRED2);

\coordinate (linkstart) at ($ (DEC1)!0.5!(DEM1) $);
\coordinate (linkend) at ($ (DEC2)!0.5!(DEM2) $);
\draw[orangearrow] (linkstart) -- (linkend);

\node[prebox,fit=(ENC1)(ENT1)(DEC1)(DEM1)] (PRETRAINBOX) {};
\node[lab,above left=0pt and -15pt of PRETRAINBOX.north,text=pretrain] {\normalsize Pre-trained};

\node[finetune,fit=(DEM2)] (FINETUNEBOX) {};
\node[lab,above right=-1pt and -25pt of FINETUNEBOX.north ,text=finetune] {\normalsize Fine-tuned};

\end{tikzpicture}
      \vspace{0em}
       \caption{}
        \label{fig:p2p_model} 
    \end{subfigure}
    \vspace{-0.5em}
    \caption{Neural compress-and-forward (CF) schemes for the diamond relay channel (DRC). (a) In the distributed scheme, each relay separately but collaboratively compresses its observation. (b) In the point-to-point (p2p) scheme, a single relay encoder and demodulator are pre-trained, and a new demodulator is fine-tuned to jointly process compressed signals coming from the two relays.
    }
    \label{fig:system_model2}
\end{figure} 


We consider two neural CF schemes based on artificial neural networks: a \textit{distributed} scheme, and a benchmark \textit{point-to-point} (p2p) scheme, where each relay compresses its observation independently, without leveraging inter-relay correlations. The former, a fully distributed compression setup, is the same as the setup in \textit{Berger--Tung coding}~\cite[Chapter~12]{network_info_theo}, where encoders compress the received signals to enable efficient transmission of correlated signals to a common decoder.

For the neural distributed CF scheme depicted in Fig.~\ref{fig:dist_model}, the relay encoders $e_{\theta_1}, e_{\theta_2}$ and their corresponding entropy coding models $q_{\zeta_1}, q_{\zeta_2}$ are parameterized by $\theta_1, \theta_2$ and $\zeta_1, \zeta_2$, respectively. The demodulator is denoted by $p_\phi$ where $\phi$ denotes its parameters. For the p2p scheme in Fig.~\ref{fig:p2p_model}, a single encoder $e_{\theta}$ and decoder $p_\xi(w \vert e_{\theta}(y_{R_1}))$ are first trained in a setting with one relay and one demodulator. In the fine-tuning phase, the encoder and entropy model parameters $\theta$ and $\zeta$ are replicated at the second relay, and a new demodulator $p_\phi(w \vert e_{\theta}(y_{R_1}), e_{\theta}(y_{R_2}))$ is trained to process both relay outputs jointly.

Note that in this p2p scheme, the shared encoder $e_{\theta}$ is unable to exploit correlations between the relay signals and therefore cannot potentially exhibit Berger--Tung like binning (i.e., grouping) in the source space. 
In contrast, as we will discuss in Section \ref{sec:sim_results}, the distributed neural CF scheme facilitates joint binning in the spirit of Berger--Tung coding, providing a good relaying strategy for the DRC with oblivious relaying~\cite{aguerri2019capacity}. The p2p, on the other hand, serves as an ablation study to assess the significance of exploiting relay correlation and enabling binning at the relay compressors.


When complex-valued modulation schemes are used, in-phase (i.e., real) and quadrature (i.e., imaginary) are compressed jointly by following the architecture in Fig. \ref{fig:system_model2}. Moreover, inspired by \cite{ozyilkan2024learning}, we also explore cases where in-phase and quadrature parts are fed into two separate encoders with distinct parameters, while still using a single demodulator. We refer to these schemes as {\em joint-IQ} and {\em split-IQ}, respectively. The compression rate of split-IQ is given by the sum of the rates achieved by in-phase and quadrature components.

It is important to highlight that the learning-based CF we adopt for the DRC builds on~\cite{ozyilkan2023learned}, where the learned compressors operate in a categorical latent space, effectively functioning as entropy-constrained vector quantizers that exploit correlation between two relay and destination signals. This is in contrast to popular class of neural compressors used for image reconstruction \cite{balle2016end, balle2018variational, minnen2018joint}, which tend to struggle with recovering discontinuous, many-to-one mappings (binning) needed to leverage decoder-only side information, as analyzed in \cite{ozyilkan2024neural}. Mirroring that rationale, we emphasize that compressors built on such transform spaces, as in popular ones~\cite{balle2016end, balle2018variational, minnen2018joint}, favor smooth transforms and thus underperform at learning binning, whereas categorical vector quantization natively supports discontinuous partitions at the latent space and performs mapping as such. As we show in Section~\ref{sec:sim_results}, our quantizers are able to emulate the {\em random binning} behavior central to Berger--Tung style coding by assigning discontinuous source regions to the same quantization index, which is then paired with entropy coding.

The goal of the CF schemes in the context of a DRC is to reduce the compression rates at two relays to satisfy link capacity constraints $R_1$ and $R_2$ (see Fig.~\ref{fig:system_model}), while maximizing the overall communication rate. Following the achievability scheme presented in~\cite[Proposition~1]{aguerri2019capacity} that provides a trade-off between the relay compression rates and end-to-end communication rate, we first set the proxy for \textit{compression rates} as follows:
\begin{align} \small
    I(Y_{R_{1}};U_{1}|U_{2}) &\leq H(U_{1}), \label{eq:r0} \\
    & \leq \mathbb{E} [ -\log_{2} q_{\zeta_{1}} (e_{\theta_{1}}(y_{R_{1}}))] \triangleq \tilde{R}_{1}, \label{eq:r1}\\
    I(Y_{R_{2}};U_{2}|U_{1}) & \leq \mathbb{E} [ -\log_{2} q_{\zeta_{2}} (e_{\theta_{2}}(y_{R_{2}}))] \triangleq \tilde{R}_{2}, \label{eq:r2}
\end{align}
where $\tilde{R}_{1} {\leq} R_{1}$ and $\tilde{R}_{2} {\leq} R_{2}$ provide operational upper bounds on each relay's compression rate. The mutual informations in~\eqref{eq:r1}  and~\eqref{eq:r2} refer to achievable CF compression rates, and the inequalities in~\eqref{eq:r1} and~\eqref{eq:r2} follow from the fact that cross-entropy is greater than or equal to entropy~\cite{network_info_theo}. Here, $\tilde{R}_i$ denotes the operational compression rate at Relay $i$, for $i \in \{1, 2\}$, where each relay employs a one-shot encoder coupled with high-order entropy coder over large blocks of the quantized signal.

Next, the \textit{communication rate} can be captured by the mutual information $I(X;U_{1}, U_{2})$, which admits the following lower bound:
\begin{align} \small
    I(X;U_{1}, U_{2}) &= H(W) - H(W | U_{1}, U_{2}), \label{eq:c_0} \\
    & \geq \log (|\mathcal{X}|) - \tilde{D}, \label{eq:c}
\end{align}
where $\tilde{D} \triangleq \mathbb{E}[ - \log (p_{\phi}(x | e_{\theta_{1}}(y_{R_{1}}), e_{\theta_{2}}(y_{R_{2}})))]$ measures the cross-entropy between the true symbol and its soft prediction. \eqref{eq:c_0} follows from the fact that $X$ is a one-to-one deterministic function of the coded symbol $W$, and \eqref{eq:c} again relies on the fact that cross-entropy is greater than equal to entropy. Since we have a fixed modulation scheme without any probabilistic shaping, we have $H(W) = H(X) = \log(|\mathcal{X}|)$. For a demodulator taking hard decisions,
\begin{align} \small
    \hat{W} = \argmax_{w \in \{ 1,\ldots,|\mathcal{X}| \}} p_{\phi}(w|e_{\theta_{1}}(y_{R_{1}}), e_{\theta_{2}}(y_{R_{2}})), \label{eq:w_hat}
\end{align}
the corresponding symbol error rate (SER) would be $ \text{SER} = P(W \neq \hat{W})$.

Building on the bounds above, the operational training objective of the neural CF scheme for the DRC
can be described by the following loss function:
\begin{align} \small
    L(\theta_{1}, \theta_{2}, \zeta_{1}, \zeta_{2}, \phi) = \left(\tilde{R}_{1} + \tilde{R}_{2} \right)  + \lambda \tilde{D}, 
\end{align}
where $\tilde{R}_1$,$\tilde{R}_2$,$\tilde{D}$ are from~\eqref{eq:r1},\eqref{eq:r2},\eqref{eq:c}, respectively, and $\lambda {>} 0$ is a trade-off parameter. The optimized models $e_{\theta_1}$,$e_{\theta_2}$, $q_{\zeta_{1}}$,$q_{\zeta_{2}}$,$p_{\phi}$ correspond to the neural compressors and entropy coders at the two relays and the joint demodulator component, respectively.

\section{Results} \label{sec:sim_results}

In this section, we present our results for the distributed neural DRC. First, we perform training under different modulation schemes BPSK, 4-PAM, 8-PAM, 4-QAM, and 16-QAM, where the symbols are equally likely, i.e., $p(x) = \frac{1}{|\mathcal{X}|}$. We define the signal power as $\mathbb{E}[|X|^{2}] \leq P$, and the SNR at Relay 1 as $\gamma_{R_{1}} = P/\sigma_{R_{1}}^{2}$, and similarly for Relay 2. We also define the average out-of-band relay rate as $R = \frac{\tilde{R}_{1} + \tilde{R}_{2}}{2}$.

We employ fully-connected neural networks with three hidden layers at both relays and the demodulator, where all of them have $128$, $256$, and $64$ neurons at each hidden layer, respectively. We use leaky rectified linear unit as the activation function and utilize the Adam optimizer~\cite{kingma2014adam}.

We evaluate our neural CF relaying schemes in terms of the trade-off between compression rates, quantified by the average rate $R$, via the bounds on $\tilde{R}_1$ in~\eqref{eq:r1} and $\tilde{R}_2$ in~\eqref{eq:r2}, and two performance metrics: (i) the end-to-end communication rate, for which we use a lower bound (serving as a conservative estimate) on $I(X;U_{1}, U_{2})$ in \eqref{eq:c} for the neural CF schemes shown in Fig.~\ref{fig:system_model2}, and (ii) the $ \text{SER} = P(W \neq \hat{W})$ (see~\eqref{eq:w_hat}).

\begin{figure}
    \centering
    \pgfplotsset{
    compat=newest,
    /pgfplots/legend image code/.code={%
        \draw[mark repeat=2,mark phase=2,#1] 
            plot coordinates {
                (0cm,0cm) 
                (0.6cm,0cm)
                (1.2cm,0cm)%
            };
    },
}
\begin{tikzpicture}
\begin{axis}[
	scale only axis,
	font=\LARGE,
	width=\textwidth,
	height=0.2*\textwidth,
	xmajorgrids=true,
	ymajorgrids=true,
	xmin=0.99,
	xmax=3.6,
        xtick={1, ..., 3},
	ymin=0.0001,
	ymax=0.04,
        ytick={0.005,0.015,...,0.04},
        y tick label style={
        /pgf/number format/.cd,
            fixed,
            fixed zerofill,
            precision=2,
        /tikz/.cd
    },
	ylabel={SER},
	legend columns=2,
	legend cell align={left},
	legend style={row sep=3pt, at={(1,1.25)}, anchor = south east, inner sep =5pt},
	]

    \addplot+[line width=2.5pt, black, mark=none, dashed] table[x=X, y=Y, col sep=tab] {
X	Y
0.9	3.412519792226882676e-02
3.6	3.412519792226882676e-02
};

    \addplot[
        line width=4pt, color=blue, mark options={scale=2,fill=blue,solid},
        mark=pentagon*]
        table [x=X, y=Y] {fig_data/multistep/ser_4pam_marginal_curve.txt};

    \addplot+[line width=2.5pt, black, mark=none, solid] table[x=X, y=Y, col sep=tab] {
X	Y
0.9	3.582436171399210403e-03
3.6	3.582436171399210403e-03
};

    \addplot[
        line width=4pt, color=magenta, mark options={scale=2,fill=magenta,solid},
        mark=square*]
        table [x=X, y=Y] {fig_data/multistep/ser_4pam_multistep_curve.txt};

    \addplot[
        line width=4pt, color=teal, mark options={scale=2,fill=teal,solid},
        mark=heart]
        table [x=X, y=Y] {fig_data/multistep/ser_4pam_relay_curve.txt};

    \legend{
            Single perfect relay,
            Neural \textit{distributed} CF (Fig.~\ref{fig:dist_model}),
            Two perfect relays,
            Neural \textit{p2p} CF (Fig.~\ref{fig:p2p_model}),
            Neural CF with single relay~\cite{ozyilkan2024learning}
    }
    
    \end{axis}
\end{tikzpicture}
    \pgfplotsset{
    compat=newest,
    /pgfplots/legend image code/.code={%
        \draw[mark repeat=2,mark phase=2,#1] 
            plot coordinates {
                (0cm,0cm) 
                (0.6cm,0cm)
                (1.2cm,0cm)%
            };
    },
}
\begin{tikzpicture}
\begin{axis}[
	scale only axis,
	font=\LARGE,
	width=\textwidth,
	height=0.2*\textwidth,
	xmajorgrids=true,
	ymajorgrids=true,
	xmin=0.99,
	xmax=3.6,
        xtick={1,...,3},
	ymin=1.85,
	ymax=2.0,
        y tick label style={
        /pgf/number format/.cd,
            fixed,
            fixed zerofill,
            precision=2,
        /tikz/.cd
    },
    ytick={1.85, 1.9, 1.95, 2.0},
	xlabel={$R$},
    ylabel={Mutual Info.},
	legend columns=1,
	legend cell align={left},
	legend style={row sep=3pt, at={(1,0)}, anchor = south east},
	]
    \addplot[
        line width=4pt, color=blue, mark options={scale=2,fill=blue,solid},
        mark=pentagon*]
        table [x=X, y=Y] {fig_data/multistep/4pam_marginal_curve.txt};

    \addplot[
        line width=4pt, color=magenta, mark options={scale=2,fill=magenta,solid},
        mark=square*]
        table [x=X, y=Y] {fig_data/multistep/4pam_multistep_curve.txt};

    \addplot[
        line width=4pt, color = teal, mark options={scale=2,fill=teal,solid},
        mark=heart]
        table [x=X, y=Y] {fig_data/multistep/4pam_relay_curve.txt};

    \addplot+[line width=3pt, black, mark=none, dashed] table[x=X, y=Y, col sep=tab] {
X	Y
0.9	1.869229999999999947e+00
3.6	1.869229999999999947e+00
};

    \addplot+[line width=3pt, black, mark=none, solid] table[x=X, y=Y, col sep=tab] {
X	Y
0.9	1.985540000000000083e+00
3.6	1.985540000000000083e+00
};

    \end{axis}
\end{tikzpicture}
    \vspace{-1.5em}
    \caption{SER and mutual information as a function of average relay rate $R = \frac{\tilde{R}_{1}+\tilde{R}_{2}}{2}$, for 4-PAM where $\gamma_{R_{1}} = \gamma_{R_{2}} = 10$dB. Solid and dashed horizontal black lines represent cases with two perfect relays (equivalent to $R_{1} \rightarrow \infty, R_{2} \rightarrow \infty$) and the single perfect relay (equivalent to $R_{1} \rightarrow \infty$, $R_{2}=0$), respectively. The green lines correspond to results from~\cite{ozyilkan2024learning}, which considered a single-relay setting with where side information is fully available at the demodulator. Therefore, the setting studied in~\cite{ozyilkan2024learning} is operationally equivalent to $R = \tilde{R}_{1}$ and $R_{2} \rightarrow \infty$. 
    Each marker on all curves represents a training run with a specific value of $\lambda$.}
    \label{fig:multistep}
\end{figure}
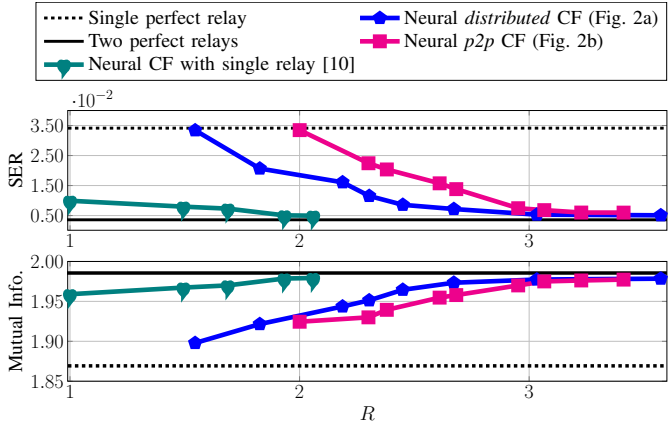

In Fig.~\ref{fig:multistep}, we compare the performance of the distributed CF scheme (Fig.~\ref{fig:dist_model}) with the p2p CF approach (Fig.~\ref{fig:p2p_model}) as a function of $R$ when 4-PAM modulation is used and SNRs at Relay 1 and Relay 2 are set as $\gamma_{R_{1}} = \gamma_{R_{2}} = 10$ dB. We also show the performance with a \textit{single perfect relay} (i.e., $R_{1}=0, R_{2} \rightarrow \infty$) and \textit{two perfect relays} (i.e., $R_{1} \rightarrow \infty, R_{2} \rightarrow \infty$) to highlight the scenarios that provide the best performance under a single relay and two relays, respectively. These cases are included because the schemes studied in this paper (Figs.~\ref{fig:system_model2}(a)-(b)), by nature, operate between these two regimes. Both the distributed and p2p schemes are also benchmarked against the performance of a neural CF scheme for the primitive relay channel with a single relay, i.e., $R = \tilde{R}_{1}$ and perfect side information at the destination, as studied in~\cite{ozyilkan2024learning}, which is operationally equivalent to $R_{2} \rightarrow \infty$. In contrast to~\cite{ozyilkan2024learning}, the DRC setup considered here involves two relays that must independently compress their noisy observations and jointly exploit correlation without any direct communication. As seen in both panels in Fig.~\ref{fig:multistep}, the distributed CF scheme outperforms the p2p scheme, particularly at low rates. We attribute this improvement to the learned one-shot joint binning behavior in the source space (visualized in Fig.~\ref{fig:4pam_visualization} and discussed later), which yields rate reduction. Moreover, as the overall rate increases, the distributed scheme approaches the performance of the single-relay setup with perfect side information at the destination, as studied in~\cite{ozyilkan2024learning}, much faster than its p2p counterpart.

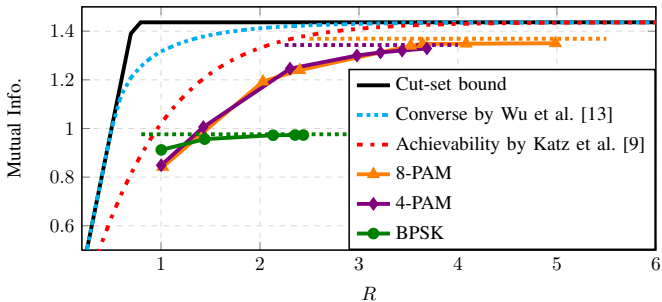
\begin{figure}[htbp]
    \centering
    \begin{tikzpicture}
    \begin{axis}[
        width=12cm,
        height=6cm,
        xlabel={$R$},
        ylabel={Mutual Info.},
        xmin=0.2,
	    xmax=6,
        xtick={1,...,6},
        ymin=0.5,
	    ymax=1.5,
        ytick={0.4, 0.6, 0.8, 1, 1.2, 1.4},
        grid=major,
        grid style={dashed,gray!30},
        thick,	
        legend columns=1,
	   legend cell align={left},
	   legend style={row sep=3pt, at={(1,0)}, anchor = south east},
    ]
    \addplot[line width=2pt,
        color=black,
        mark=None]
        table [x=X, y=Y] {fig_data/real_marginal/cutset_curve.txt};

    \addplot[line width=2pt,
        color=cyan,
        mark=None, dashdotdotted]
        table [x=X, y=Y] {fig_data/real_marginal/new_bound_curve.txt};

    \addplot[line width=2pt,
        color=red,
        mark=None, loosely dashdotted]
        table [x=X, y=Y] {fig_data/real_marginal/shlomo_curve.txt};

    \addplot[line width=2pt,
        color=orange,
        mark=triangle*]
        table [x=X, y=Y] {fig_data/real_marginal/8pam_curve.txt};

    \addplot[line width=2pt,
        color=violet,
        mark=diamond*]
        table [x=X, y=Y] {fig_data/real_marginal/4pam_curve.txt};

    \addplot[line width=2pt,
        color=green,
        mark=*]
        table [x=X, y=Y] {fig_data/real_marginal/bpsk_curve.txt};

    \addplot+[line width=2pt, orange, mark=none, dotted] table[x=X, y=Y, col sep=tab] {
X	Y
2.5	1.368773166718259882e+00
5.5	1.368773166718259882e+00
};

    \addplot+[line width=2pt, green, mark=none, dotted] table[x=X, y=Y, col sep=tab] {
X	Y
0.8	9.761772335954986834e-01
3	9.761772335954986834e-01
};

    \addplot+[line width=2pt, violet, mark=none, dotted] table[x=X, y=Y, col sep=tab] {
X	Y
2.25	1.343122877358368772e+00
4	1.343122877358368772e+00
};

    \legend{
        Cut-set bound,
        Converse by Wu et al.~\cite{ayfer2019new},
        Achievability by Katz et al.~\cite{katz2024gaussian},
        8-PAM,        
        4-PAM,
        BPSK
    }
    
    \end{axis}
\end{tikzpicture}
    \caption{Mutual information for the distributed scheme (Fig. ~\ref{fig:dist_model}) under BPSK, 4-PAM, and 8-PAM modulations with $\gamma_{R_{1}} = \gamma_{R_{2}} = 5$ dB. For the bounds obtained from~\cite{ayfer2019new, katz2024gaussian}, we choose the rates for both relays as $R = R_{1} =R_{2}$. Dashed horizontal lines represent performance of two perfect relays (i.e., $R_{1} \rightarrow \infty$, $R_{2} \rightarrow \infty$) for the respective curves, similar to Fig.~\ref{fig:multistep}.}
    \label{fig:fig1}
\end{figure}

In Fig.~\ref{fig:fig1} mutual information performances of BPSK, 4-PAM, and 8-PAM for the distributed scheme are provided for different rates $R$ with $\gamma_{R_{1}} = \gamma_{R_{2}} = 5$ dB, where the dashed  horizontal lines indicate the mutual information with two perfect relays in Fig.~\ref{fig:multistep}, i.e., $R_{1} \rightarrow \infty$, $R_{2} \rightarrow \infty$, under the corresponding modulation scheme. For the asymptotic achievability and converse baselines, we have used results in~\cite{ayfer2019new, katz2024gaussian}, respectively. Results show that the distributed neural CF scheme in each modulation scheme can reach their respective asymptotic capacity and can get closer to the theoretical bounds that assume Gaussian input as the modulation order increases \cite{ayfer2019new}. 

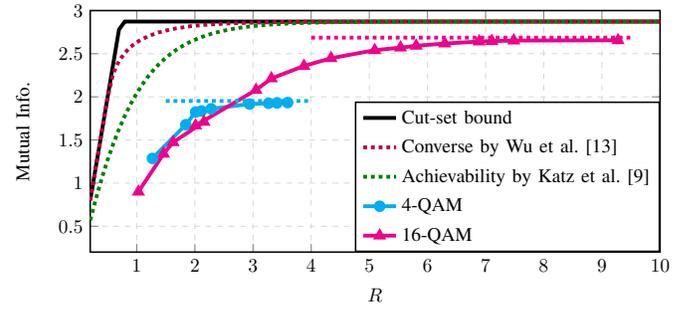
\begin{figure}[htbp]
    \centering
    \begin{tikzpicture}
    \begin{axis}[
        width=12cm,
        height=6cm,
        xlabel={$R$},
        ylabel={Mutual Info.},
        xmin=0.2,
	    xmax=10,
        xtick={1,...,10},
        ymin=0.2,
	    ymax=3,
        ytick={0.5,1,1.5,2,2.5,3},
        grid=major,
        grid style={dashed,gray!30},
        thick,	
        legend columns=1,
	   legend cell align={left},
	   legend style={row sep=3pt, at={(1,0)}, anchor = south east},
    ]
    \addplot[line width=2pt,
        color=black,
        mark=None]
        table [x=X, y=Y] {fig_data/complex_marginal/complex_cutset_curve.txt};

    \addplot[line width=2pt,
        color=purple,
        mark=None, dotted]
        table [x=X, y=Y] {fig_data/complex_marginal/complex_new_bound_curve.txt};

    \addplot[line width=2pt,
        color=green,
        mark=None, dotted]
        table [x=X, y=Y] {fig_data/complex_marginal/complex_shlomo_curve.txt};

    \addplot[line width=2pt,
        color=cyan,
        mark=*]
        table [x=X, y=Y] {fig_data/complex_marginal/4qam_curve.txt};

    \addplot[line width=2pt,
        color=magenta,
        mark=triangle*]
        table [x=X, y=Y] {fig_data/complex_marginal/16qam_curve.txt};

    \addplot+[line width=2pt, cyan, mark=none, dotted] table[x=X, y=Y, col sep=tab] {
X	Y
1.5	1.952354467190997589e+00
4	1.952354467190997589e+00
};

    \addplot+[line width=2pt, magenta, mark=none, dotted] table[x=X, y=Y, col sep=tab] {
X	Y
4	2.686245754716737100e+00
9.5	2.686245754716737100e+00
};

\legend{
        Cut-set bound,
        Converse by Wu et al.~\cite{ayfer2019new},
        Achievability by Katz et al.~\cite{katz2024gaussian},
        4-QAM,
        16-QAM
    }

    \end{axis}
\end{tikzpicture}
    \caption{Mutual information for the distributed scheme (Fig.\ref{fig:dist_model}) under 4-QAM and 16-QAM with $\gamma_{R_{1}} {=} \gamma_{R_{2}} {=} 5$dB. For the bounds obtained from~\cite{ayfer2019new, katz2024gaussian}, we choose the rates for the relays as $R {=} R_{1} {=} R_{2}$. Dashed horizontal lines represent the case of perfect relays, i.e., $R_{1} {\rightarrow} \infty$, $R_{2} {\rightarrow} \infty$ under its corresponding modulation scheme.}
    \label{fig:4qam_16qam}
\end{figure}

\begin{figure}
    \centering
    \begin{subfigure}[t]{0.46\columnwidth}
        \centering
        \includegraphics[width=\columnwidth]{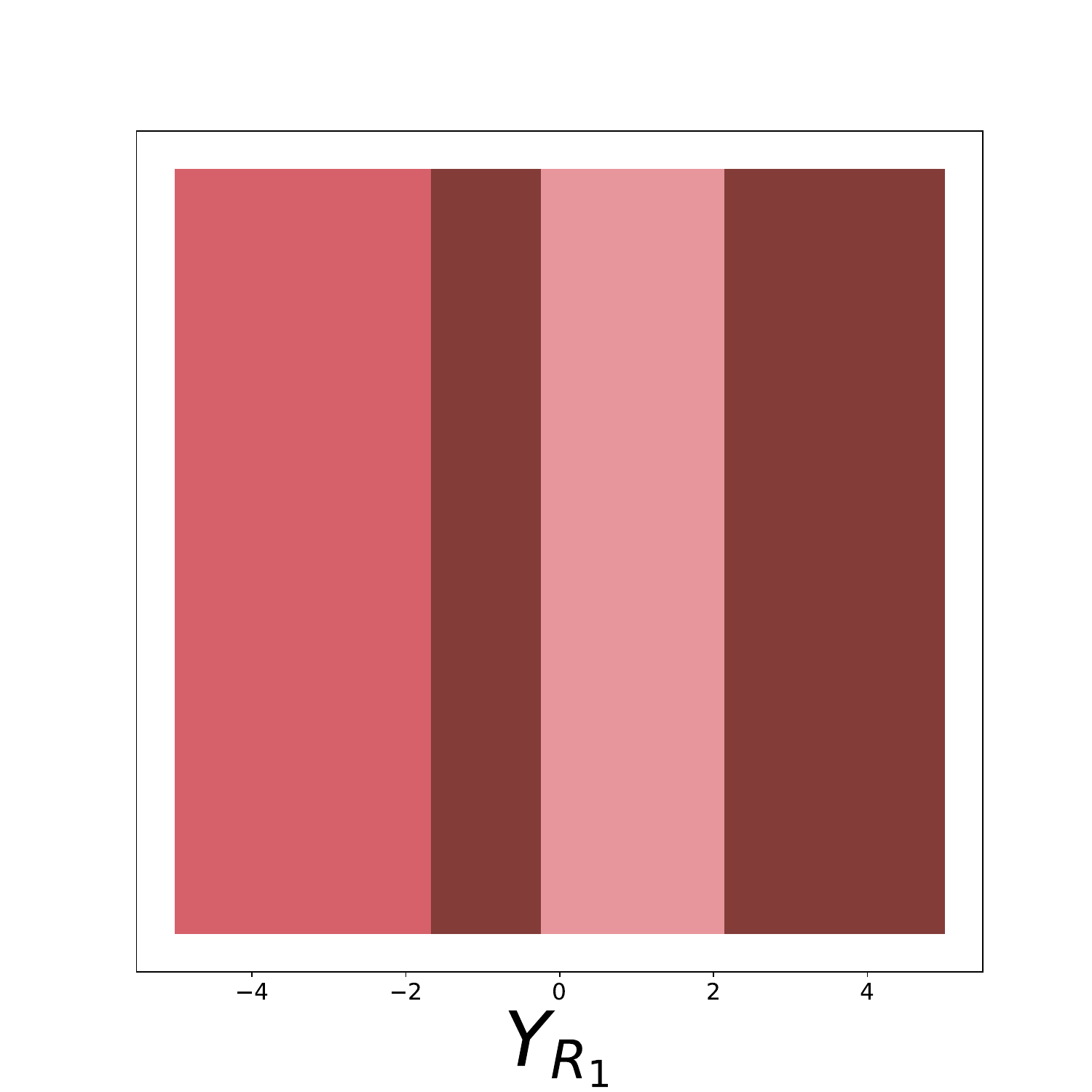}
        \vspace{-1.5em}
        \caption{}
        \label{fig:enc1}
    \end{subfigure}
    \begin{subfigure}[t]{0.46\columnwidth}
        \centering
        \includegraphics[width=\columnwidth]{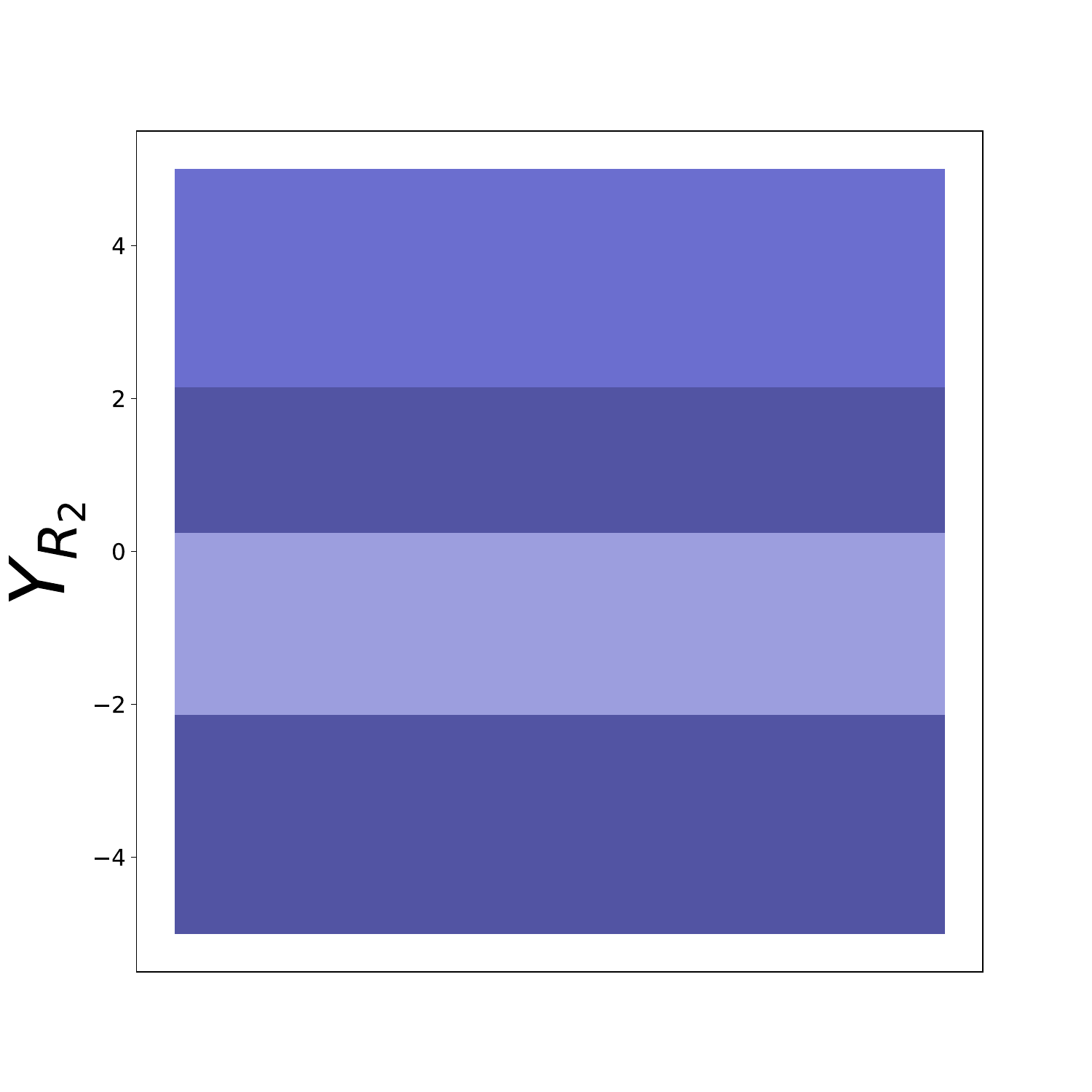}
        \vspace{-1.5em}
        \caption{}
        \label{fig:enc2}
    \end{subfigure}


    \begin{subfigure}[t]{0.9\columnwidth}
        \centering
        \includegraphics[width=\columnwidth]{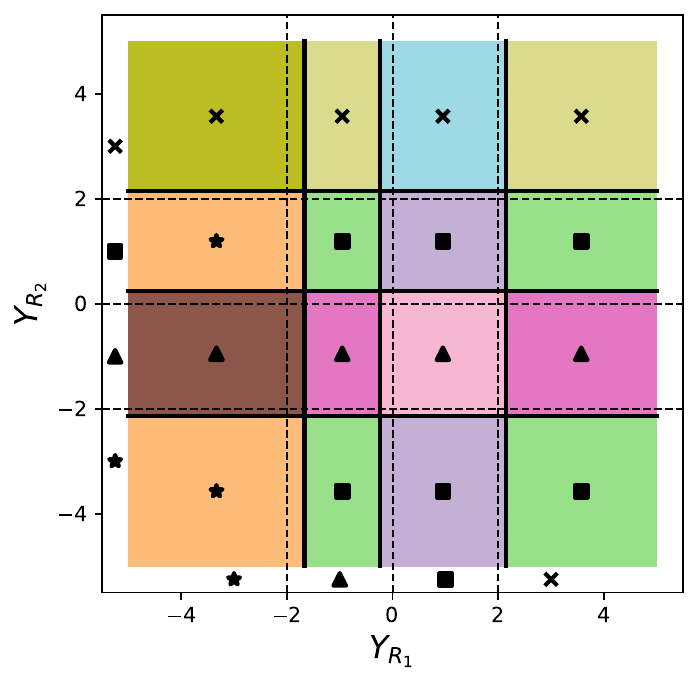}
        \caption{}
        \label{fig:demod}
    \end{subfigure}
    \caption{Visualization (best viewed in color) of learned distributed encoders, $e_{\theta_1}\left(Y_{R_1}\right)$ in (a) and $e_{\theta_2}\left(Y_{R_2}\right)$ in (b), and demodulator decisions in (c) for 4-PAM modulation when $R \approx 1.50$ and $\gamma_{R_{1}} = \gamma_{R_{2}} = 10$dB. Different colors for encoders represent distinct quantization indices $e_{\theta_{1}}(Y_{R_{1}})$ and $e_{\theta_{2}}(Y_{R_{2}})$, while the colors in the decision regions of the demodulator correspond to unique combinations of quantization indices received from the two relays. Vertical and horizontal lines in (a) and (b) indicate the decision boundaries of the relay encoders, and markers in (c) represent the hard decisions made at the demodulator. The transmitted symbols by each relay are also shown near the axis for reference.}
    \label{fig:4pam_visualization}
\end{figure}

In Fig. \ref{fig:4qam_16qam}, we provide the performance of 4-QAM and 16-QAM under the distributed scheme for various $R$ values when $\gamma_{R_{1}} = \gamma_{R_{2}} = 5$ dB. The dashed horizontal lines for each modulation scheme indicate the case with two perfect relays, i.e., $R_{1} \rightarrow \infty$, $R_{2} \rightarrow \infty$. The mutual information values are benchmarked against the asymptotic achievability and converse baselines, respectively from~\cite{ayfer2019new, katz2024gaussian}, as well as the cut-set bound. For 4-QAM, the joint-IQ method is employed throughout the simulations, whereas for 16-QAM, the convex hull of split-IQ and joint-IQ methods is considered, since the best performing scheme varies with $R$. 

In Fig.~\ref{fig:4pam_visualization}, we visualize the 4-PAM quantization regions of the learned encoders and decision regions of the demodulator, for $\gamma_{R_{1}} {=} \gamma_{R_{2}} {=} 10$ dB and $R {\approx} 1.50$. Figs.~\ref{fig:enc1} and~\ref{fig:enc2} show the encoder mappings $e_\theta(Y_{R_1})$ and $e_\theta(Y_{R_2})$ learned at the two relays, where colors represent the transmitted quantization indices and the grid lines indicate the quantization boundaries. Fig.~\ref{fig:demod} displays the decision regions at the destination, with horizontal and vertical axes given by $Y_{R_1}$ and $Y_{R_2}$, respectively. The color-coded regions in Figs.~\ref{fig:4pam_visualization}(a)–(c) reveal binning behavior, as non-adjacent intervals are mapped to the same index (color). In Fig.~\ref{fig:demod}, the lines represent the hard decision boundaries for each combination of quantization indices received from the two relays, and the markers denote the demodulated messages $\hat{W}$ as in~\eqref{eq:w_hat}. As observed, these boundaries are not only shifted relative to the midpoints between PAM symbols, which would be optimal thresholds without relaying, but are also more finely partitioned. 

This figure also reflects how the demodulator at the destination learns to make decisions over possible combinations of received quantized indices. For instance, when the square symbol is transmitted, the encoders are likely to produce index light red from Relay 1 and index light purple from Relay 2. In this case, the corresponding decision region for the square symbol at the destination becomes larger than those of other symbols, demonstrating how the learned demodulator adapts the likelihood $p_{\phi}(w \vert e_{\theta_1}(y_{R_1}), e_{\theta_2}(y_{R_2}))$ based on the received indices from the two relays. Moreover, we observe that the joint combinations of indices from both encoders can result in the same combination being assigned to multiple disjoint regions. For example, the combination represented by the orange color in Fig.~\ref{fig:demod}, which appears in two nonadjacent regions. 

To illustrate this behavior more precisely, we consider the following example. Even when the signal $Y_{R_2}$ lies within a region typically associated with the star symbol, the demodulator may instead assign the square symbol. This occurs due to the shared light green index across the four nonadjacent regions, as shown in Fig.~\ref{fig:demod}. Such a pattern suggests that the encoders prioritize finer quantization around the origin, where multiple decision boundaries cluster, while tolerating overlap in the off-center regions. This trade-off reduces the SER by focusing on high error probability areas and simultaneously leverages binning by assigning the same index to nonadjacent regions, thereby reducing the required compression rate. 

Notably, such a compress-bin strategy does not emerge (not shown) in the p2p scheme (Fig.~\ref{fig:p2p_model}), where the encoders, by nature, cannot exploit any inter-relay correlation. As a result, the p2p scheme tends to require a higher rate to achieve a similar SER level with the distributed one, particularly in the low-rate regime, as seen in Fig.~\ref{fig:multistep}.

\section{Conclusion}
In this paper, we have extended the application of neural CF scheme to the DRC setup, where two separated relays compress their noisy observations and forward them to the destination for joint decoding. To this end, the relay compressors and the demodulator were parameterized by fully connected neural networks and trained end-to-end. Simulation results demonstrate that the proposed neural distributed CF scheme consistently outperforms the benchmark {\em p2p} scheme, while approaching the asymptotic behavior and operating close to theoretical limits as the average relay rate increases. We have evaluated performance across various modulation schemes, both real and complex, and provided an explanation on how the distributed neural CF architecture induces decision regions at the destination that exhibit joint binning, resulting in reduced compression rate. 

As a future work, we plan to investigate robustness under heterogeneous relay conditions, such as unequal SNRs or asymmetric rate constraints. Another promising direction is to generalize the framework to multi-source networks, where relays must compress signals coming from multiple transmitters.


\bibliography{references}
\bibliographystyle{ieeetr}

\end{document}